\documentclass[twocolumn,superscriptaddress,prl,floatfix]{revtex4}
\usepackage[]{graphicx}
\usepackage[usenames,dvipsnames]{color}
\usepackage{soul}
\usepackage{bm}

\newcommand{\cubbio}{Bi$_2$CuO$_4$}

\newcommand{\pr}{%
        ^\prime}
    
\newcommand{\svek}{%
        \mathbf}
\newcommand{\vek}[1]{%
        \hbox{\textbf #1}}

\begin{document}

\title{Strongly correlated double Dirac fermions}

\author{Domenico Di Sante}
\affiliation{Institut f\"{u}r Theoretische Physik und Astrophysik,
  Universit\"{a}t W\"{u}rzburg, Am Hubland Campus S\"{u}d,
  W\"{u}rzburg 97074, Germany}

\author{Andreas Hausoel}
\affiliation{Institut f\"{u}r Theoretische Physik und Astrophysik,
  Universit\"{a}t W\"{u}rzburg, Am Hubland Campus S\"{u}d,
  W\"{u}rzburg 97074, Germany}

\author{Paolo Barone}
\affiliation{Consiglio Nazionale delle Ricerche (CNR-SPIN), Via Vetoio, L’Aquila, Italy}

\author{Jan M. Tomczak}
\affiliation{Institute of Solid State Physics, Vienna University of Technology, A-1040 Vienna, Austria}

\author{Giorgio Sangiovanni}
\affiliation{Institut f\"{u}r Theoretische Physik und Astrophysik,
  Universit\"{a}t W\"{u}rzburg, Am Hubland Campus S\"{u}d,
  W\"{u}rzburg 97074, Germany}

\author{Ronny Thomale}
\affiliation{Institut f\"{u}r Theoretische Physik und Astrophysik,
  Universit\"{a}t W\"{u}rzburg, Am Hubland Campus S\"{u}d,
  W\"{u}rzburg 97074, Germany}

\date{\today}

\begin{abstract}
Double Dirac fermions have recently been identified as possible
quasiparticles hosted by three-dimensional crystals with particular
non-symmorphic point group symmetries. Applying a combined approach
of ab initio methods and dynamical mean field theory, we investigate how interactions and
double Dirac band topology conspire to form the electronic quantum
state of \cubbio. We derive a downfolded eight-band model of the
pristine material at low
energies around the Fermi level.
By tuning the model parameters from the free band structure
to the realistic strongly correlated regime, we find a persistence of
the double Dirac dispersion until its constituting time reveral
symmetry is broken due to the onset of magnetic ordering at the Mott
transition. We analyze pressure as a promising route to realize a
double-Dirac metal in {\cubbio}.
\end{abstract}


\maketitle


{\it Introduction.} Electrons in solids witness a reduced spatial
symmetry in comparison to the spacetime continuum. While the high-energy
perspective constrains us to Majorana, Weyl, and Dirac
fermions in accordance with the inhomogeneous Lorentz (or Poincar\'e)
group, electronic quasiparticles in solids at low energies can display
emergent fermionic behaviour within and even beyond this classification~\cite{charly,berni}. Graphene
constitutes one of the most prominent material discoveries where
Dirac-type quasiparticles have
been realized~\cite{neto}. The current rise of Weyl semimetals~\cite{wan}, along
with Majorana quasiparticles in superconducting
heterostructures~\cite{nick,fu}, complements this evolution in contemporary condensed
matter physics. 

Recently, Wieder {\it et al.}~\cite{charly} have brought up the possibility to
realize double Dirac quasiparticles in certain 3D crystals with specific
non-symmorphic point group symmetry. This was
followed up on by a systematic analysis of all double space groups (SGs)
accounting for $S=1/2$ electrons in spin-orbit coupled crystals with
time reversal symmetry~\cite{berni}. In particular, SG
130 and 135 were found to establish eminently suited ground for
generic double Dirac fermions protected by point group symmetry. Among
all material candidates for SG 130, it is was already realized
in Ref.~\onlinecite{berni} that {\cubbio}  might be a prime candidate
due to its filling-enforced semimetallicity~\cite{michi}, nurturing
the hope to observe double Dirac fermions close to the Fermi
level. All such band structure classifications, however, always need to be
extended to account for the role of interactions in the
material, which turn out to be of vital relevance in \cubbio. Most of
the topological band properties, even the
metallic ones, display a certain
degree of persistence against weak interactions as long as those do not
break any protecting symmetry. Interaction-induced instabilities, however, do
change the symmetry class of the quantum state, possibly affecting the
whole range of constituting symmetries including charge conservation,
time reversal, and point group operations.  

In this Letter, we study strong interaction effects of double
Dirac fermions, analyzing the band structure
properties and correlation effects in {\cubbio}. From density
functional theory, we distill an effective eight-band tight-binding Hubbard model which is
dominated by the $d_{x^2-y^2}$ orbital of the four Cu atoms in the
unit cell. Spin-orbit coupling (SOC) is found to be weak because the heavy
atoms of the compound do not significantly contribute to the
low-energy density of states. As expected, the double Dirac dispersion
is located close to the Fermi energy, as identified by an 8-fold
band degeneracy at the A point. We quantify the strength of
electron-electron interactions by means of the
constrained random phase approximation (cRPA). 
By employing a combined approach of density functional theory
(DFT) and dynamical mean field theory (DMFT), we determine the critical
value for the Mott transition at ambient pressure to be
$U_{\text{Mott}}\approx 0.85$eV. This value is approximately equal to the $d_{x^2-y^2}$ bandwidth, and substantially smaller
than $U_{\text{cRPA}}\approx 1.58$eV, suggesting that {\cubbio} at
ambient pressure parametrically locates itself deep in the Heisenberg
limit of the Mott regime. Below the N\'eel
temperature T$_N$
the magnetic state displays an AFM-C type collinear order of the four
local Cu magnetic moments.
In the stochiometric compound at pristine
conditions, magnetism thus dominates the electronic state, rendering
the double Dirac fermions inaccessible. We explore the
effect of hydrostatic pressure to drive {\cubbio} into a metallic phase where the
double Dirac fermions emerge. 
We find the bandstructure formation and correlation effects of {\cubbio } to be highly
sensitive to pressure.  Starting from ambient pressure, we find an initial change of the
magnetic ordering from AFM-C to AFM-G type, along with a reduced
on-site Coulomb repulsion. This could pave the way towards a high-pressure
double Dirac metal. 

 \begin{figure}[!t]
 \centering
 \includegraphics[width=\columnwidth,angle=0,clip=true]{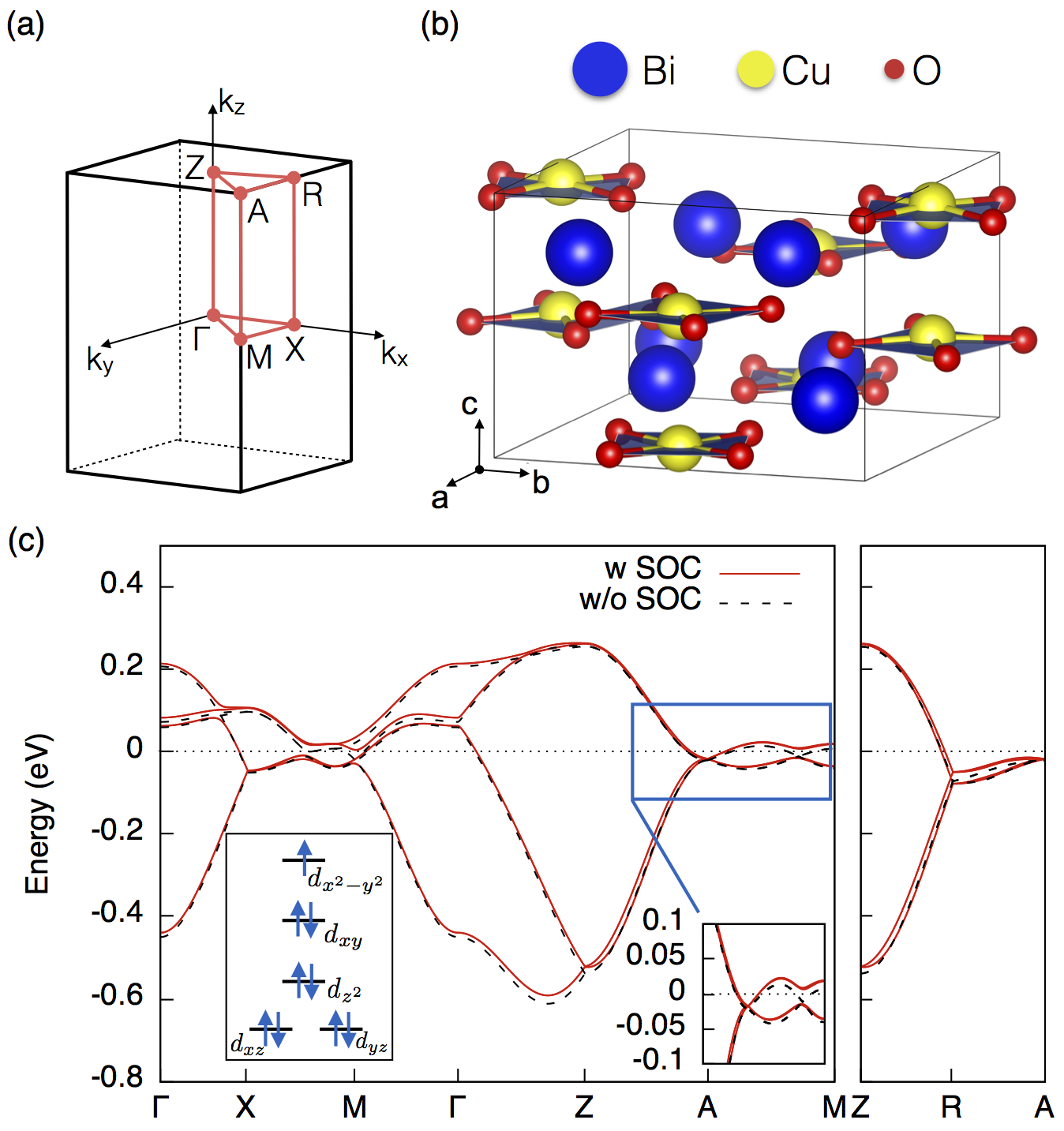}
 \caption{
 (Color online) (a) Tetragonal Brillouin zone of SG 130 P4/ncc with corresponding high-symmetry points.
 (b) Crystal structure of {\cubbio}. Atom colors: Bi = blue, Cu = yellow, O = red. Blue plaquettes
 highlight the square-planar CuO$_4$ coordination. (c) DFT bandstructure in the metallic phase with
 (red solid line) and without (black dashed line) SOC. The left inset is a schematic representation of
 the square-planar crystal field and orbital filling for Cu-type d$^9$. The right inset shows a zoom around
 the eight-fold degenerate A point, which comes along a characteristic four-fold
 degneracy along the R-Z line.}
 \label{fig1}
 \end{figure}

\begin{table}[!t]
\centering
\caption{Hopping integrals in meV between the d$_{x^2-y^2}$-type Wannier functions
on different Cu atoms for different pressure P [GPa] with ambient
pressure P $=0$ GPa. The first row takes over the notation
introduced in Ref.~\cite{Janson}, while t$_{i,j}$ in the second row is transfer
integral between Cu-$i$ and Cu-$j$ site ($i,j = 1...4$) as in Fig. \ref{fig2}. When a $j$ index has a superscript
$\pm 1$ it refers to a Cu-$j$ atom in the [00$\pm 1$] unit cell.  
$U_{\text{cRPA}}$ (eV) is the cRPA value for the effective onsite Coulomb interaction.
The $P=0$ hopping values in parenthesis refer to Ref. \cite{Janson}.}
\label{tab1}
\begin{tabular}{p{0.8cm}p{0.8cm}p{0.8cm}p{0.8cm}p{0.8cm}p{0.8cm}p{0.8cm}p{0.8cm}}
\hline\hline
P & t$^{AB}_{1u}$ &  t$^{AB}_{1d}$ &  t$^{AB}_{2u}$  &  t$^{AB}_{2d}$           &   t$^A_1$    & t$^A_2$ & $U_{\text{cRPA}}$ \\
  & t$_{1,2}$    &  t$_{4,2}$    &  t$_{1,3}$     &  t$_{1,2^{-1}}$  &   t$_{1,4}$  &  t$_{1,1^{1}}$  & \\
\hline
0 &  66  &  40  &    5 &  -24  &  36  &  -20  &  1.58   \\
  & (74) & (36) &    - & (-40) & (21) & (-18) &         \\
30&  80  &  21  &    6 &  -20  &  73  &  -16  &  1.45   \\
50&  88  &  13  &    2 &  -13  &  65  &  -7   &  1.36   \\
\hline
\hline
\end{tabular}
\end{table}

 \begin{figure}[!h]
 \centering
 \includegraphics[width=\columnwidth,angle=0,clip=true]{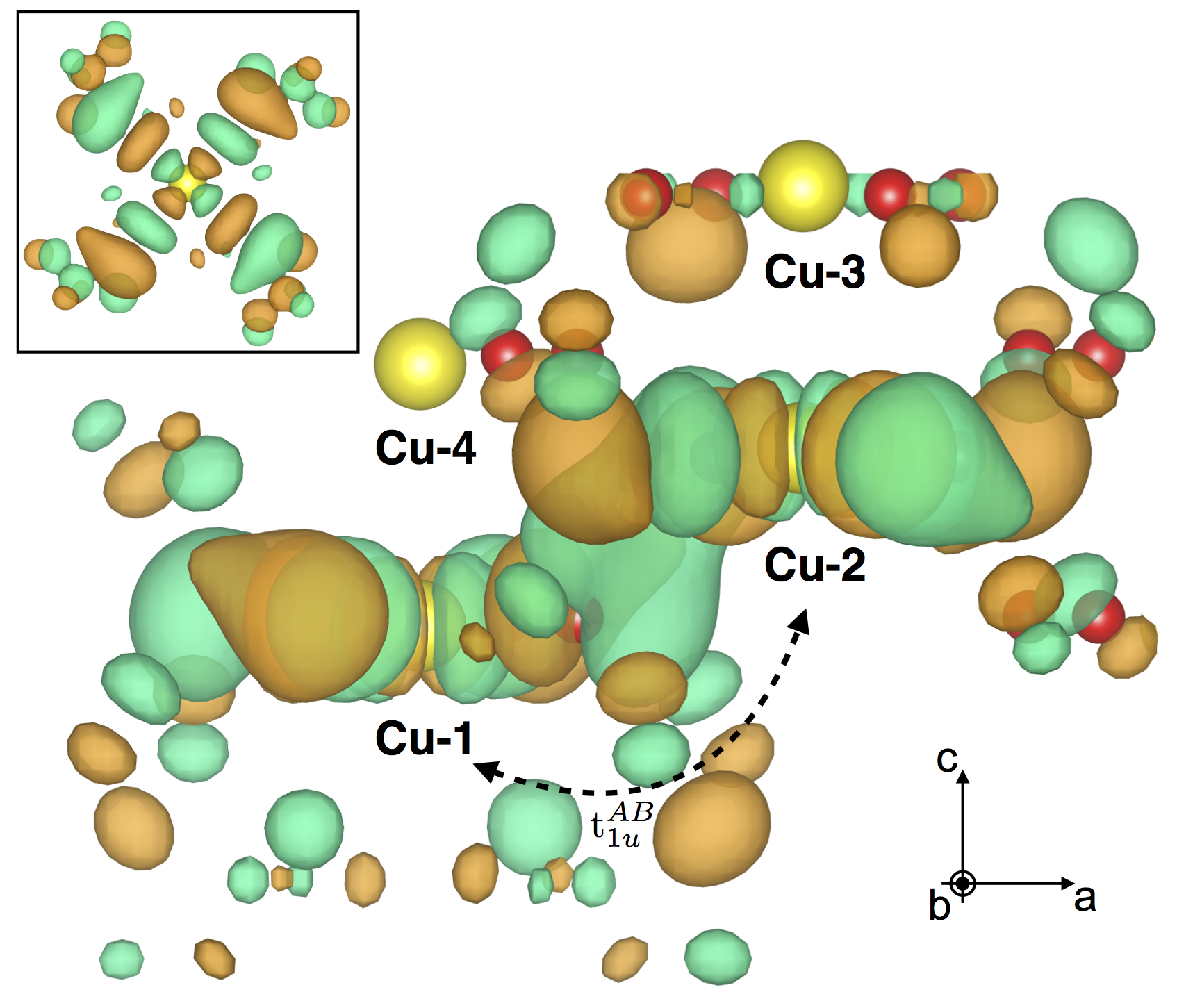}
 \caption{
 (Color online) Side view of the d$_{x^2-y^2}$-type MLWFs belonging to two neighbouring Cu-chains, responsible for the transfer integral t$^{AB}_{1u}$ (Tab. \ref{tab1}).
 Brown (green) color refers to positive (negative) values of the MLWF.
 The individual MLWFs are not symmetric with respect to the CuO$_4$ plaquette plane.
 This is due to a small off-centering of the Cu atoms, in opposite directions for Cu-1 and Cu-2.
 The inset shows a top view on the Cu-1 d$_{x^2-y^2}$-type MLWF.}
 \label{fig2}
 \end{figure}
 
 \begin{figure*}[!t]
 \centering
 \includegraphics[width=\textwidth,angle=0,clip=true]{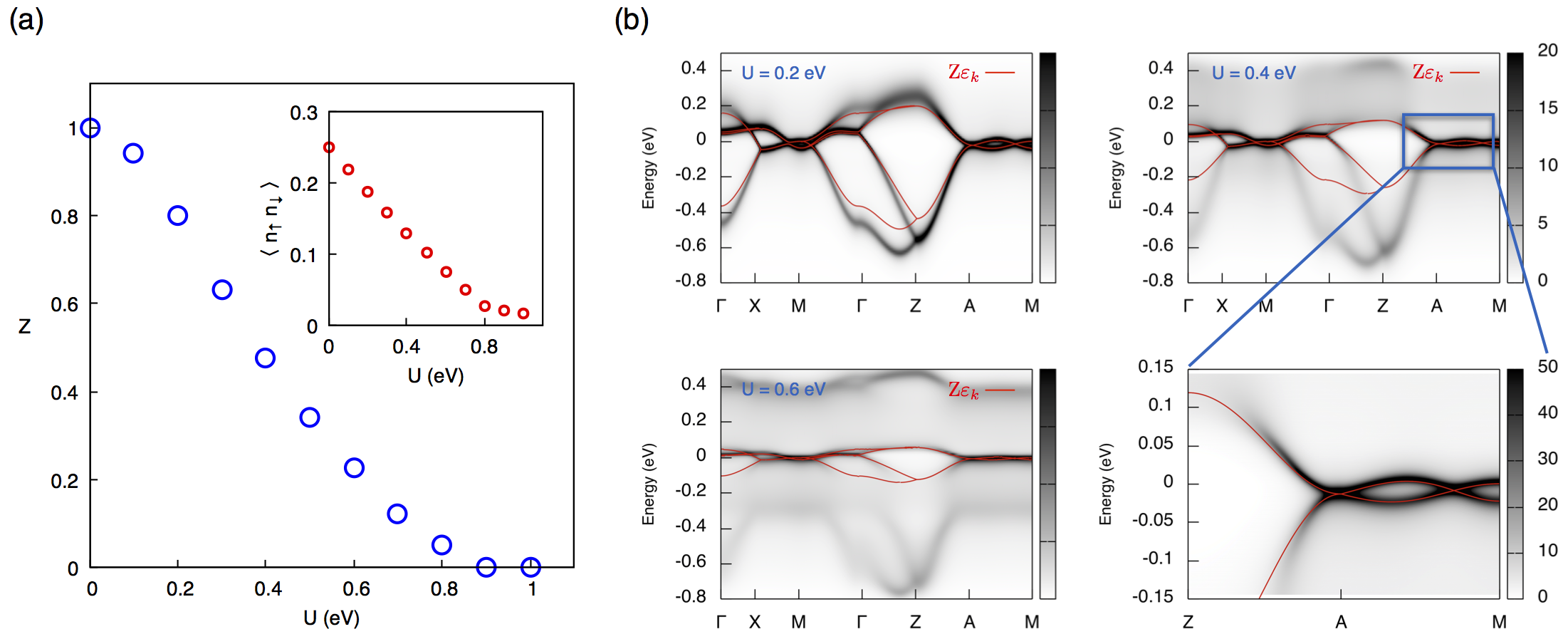}
 \caption{
 (Color online) (a) DMFT quasiparticle weight $Z$ (main panel) and double occupancies
 (inset) as a function of $U$ in the paramagnetic phase for a
 temperature $\beta = 100$ eV$^{-1}$. The Mott transition occurs at $U_{\text{Mott}} \approx 0.85$ eV. (b) Momentum-resolved DMFT spectral function A$(k,\omega)$
 for $U$ = 0.2, 0.4, and 0.6 eV. Red lines denote the DFT band
 structure renormalized by the DMFT quasiparticle weight $Z$. 
 For $U = 0.4$ eV, the zoom shows the preservation of the eight-fold
 degenerate A point along with an already signficant band renormalization.}
 \label{fig3}
 \end{figure*}

{\it Band structure and effective tight-binding model.} {\cubbio}
crystallizes in the tetragonal spage group P4/ncc (SG 130) (Fig.
\ref{fig1}(a)) \cite{GarciaMunoz}. Four inequivalent Cu atoms in the unit cell share a
square-planar CuO$_4$ coordination, stacked along the out-of-plane
direction and intercalated by Bi atoms, as depicted in Fig.~\ref{fig1}(b). 
The Cu-O hybridization is quite strong, leading to a
mixed $d$-$p$ character of the band structure of the occupied states~\cite{Supp}. 
The Bi-$p$ manifold only starts $\approx 1.5$ eV above the Fermi
level, suggesting that the low energy model is given by Cu and O electronic states. The square-planar crystal field
is characterized by a higher-in-energy $d_{x^2-y^2}$ orbital experiencing a head-to-head interaction
with the O-$p$. The level separation is schematically drawn in the left inset
of Fig.~\ref{fig1}(c). Cu is nominally in the oxidation state 2+ (i.e. $d^9$ configuration). 
For this reason, a half-filled $d_{x^2-y^2}$-derived single spin-degenerate band per Cu atom in the unit cell is expected to dominate at low energies. 
Fig.~\ref{fig1}(c) displays the band dispersion of the
electronic states around the Fermi level. Due to the dominant Cu
orbital, the
effect of SOC is almost negligible, only inducing 
small spin splittings. We estimate the energy scale of SOC to be $\approx
20$ meV. Despite its small scale, however, SOC
is a crucial ingredient for the symmetries in the double SG
representation that ensure the existence of a double Dirac fermion with
linear dispersion at the Brillouin zone corner $\text{A}=(\pi,\pi,\pi)$. As the band filling
is given by an odd multiple of 4 (180 = 8$\times$22+4), the double Dirac fermion in
{\cubbio} is located almost at the Fermi level, as visible in the right
inset of Fig.~\ref{fig1}(c). 

While the aforementioned electronic structure details, as well as the magnetic properties
at ambient pressure, have already been analyzed previously~\cite{Janson,Parmigiani},
a closer inspection of the hybridization profile and the Coulomb matrix elements in terms of ab-initio derived maximally localized 
Wannier functions (MLWFs) \cite{Marzari} is indispensible to further
analyze the fate of double Dirac fermions in the presence of interactions.
The relatively large extension of the MLWFs is a consequence of the strong Cu-O hybridization. 
Their lobes indeed stretch out over the O-atoms with a clockwise winding shape, as shown in the top view of Fig.~\ref{fig2}. 
Our analysis elucidates the origin of the strong three-dimensional
character of {\cubbio}, as the MLWFs extend along the out-of-plane 
direction in an asymmetric manner, which is a consequence of buckled CuO$_4$ plaquettes. 
Moreover, two $d_{x^2-y^2}$-type MLWFs localized on different chains have a strong overlap, even though
they do not belong to the same $ab-$plane. This is illustrated by the
t$^{AB}_{1u}$ transfer integral shown in Fig.~\ref{fig2} and Tab.~\ref{tab1}. 
Coherent electron hopping is also enabled between neighbouring
plaquettes along the normal $c-$direction. This process is accounted for by the transfer integral
t$^{A}_1$, which is indeed comparable with the hoppings t$^{AB}_{1u}$
and t$^{AB}_{1d}$ that involve a
significant in-plane component. As further discussed below, the
hoppings turn out to be rather sensitive to
pressure, and even impose a change of the magnetic ordering pattern,
which is relevant to classifying the possible topological
character of symmetry-broken phases in {\cubbio}~\cite{Schnyder}. 

{\it Mott transition and spectral function.} Having obtained an effective
band structure description of {\cubbio} close to $E_{\text{F}}$ in terms of a half-filled single band per Cu atom,
we focus on the double Dirac point at the A point. 
In particular, we analyze its evolution as a function of interaction
strength, starting from the free electron limit
up to the realistic cRPA value for the electron-electron interaction
$U$ \cite{note_UcRPA}.  {\cubbio} is parametrically located within the applicability bounds of
DFT+DMFT, partly because of its three-dimensional character.
Fig. \ref{fig3}(a) displays the 
quasiparticle weight 
$Z\!=\!(1\!-\!\partial\text{Im} \Sigma(\omega)/\partial\omega |_{\omega \rightarrow 0}\!)^{-1}$.
Before reaching $U_{\text{cRPA}}$, we encounter 
a Mott-type metal-to-insulator transition (MIT) at $U_{\text{Mott}} \approx 0.85$ eV.
At the Mott transition, the quasiparticle weight is suppressed by interactions, and the fraction of 
doubly-occupied Cu-sites $\langle n_{\uparrow}n_{\downarrow}\rangle$
reduces from the free-particle value of $1/4$ towards zero (inset in Fig.~\ref{fig3}(a)). 
At intermediate $U$, the spectral function,
as depicted in Fig.~\ref{fig3}(b), is well described by the DFT single particle
band structure renormalized in terms of bandwidth by the quasiparticle
weight $Z$, along with the appearance of incoherent lower and upper Hubbard
bands. Since the double Dirac fermion is located close to the Fermi
level, it contributes to the quasiparticle peak for interaction
strengths below the Mott transition. Even though the double Dirac velocities are strongly reduced by electron-electron interactions, the
eightfold-degeneracy of the double Dirac point is preserved up to $U_{\text{Mott}}$. Within DMFT, 
as long as we consider the paramagnetic phase which does not allow the
system to undergo a magnetic phase transition, the double
Dirac point does not get destroyed for larger $U$. Yet, its weight
gets damped by electron-electron scattering,
and transferred towards the
high-energy (Hubbard) bands, alongside with the whole low-energy spectral
weight.

\begin{table}[!t]
\centering
\caption{DFT+U total energy differences (meV/unit cell) for the
  magnetic patterns at ambient pressure and at 30 GPa. The lowest-lying AFM
  type state is chosen as reference energy. Non-magnetic (NM) and
  ferromagnetic (FM) configurations are higher up in energy than the
  AFM states.}
\label{tab2}
\begin{tabular}{p{1.3cm}p{1.3cm}p{0.8cm}p{0.8cm}p{1.1cm}p{1.15cm}p{1.15cm}}
\hline\hline
 P (GPa) & $U$ (eV) & NM &  FM  & AFM-C &  AFM-G &  AFM-A \\
\hline
0  &      &     &      &       &         &      \\
   & 0.0  & 442 &  88  &    0  &    2.4  &   54 \\
   & 0.5  & 555 &  77  &    0  &    2.2  &   49 \\
   & 1.58 & 812 &  60  &    0  &    1.8  &   41 \\
   & 2.1  & 940 &  54  &    0  &    1.6  &   37 \\
30 &      &     &      &       &         &      \\
   & 1.45 & 636 &  97  &   28  &    0    &   65 \\  
\hline
\hline
\end{tabular}
\end{table}

 \begin{figure}[!t]
 \centering
 \includegraphics[width=\columnwidth,angle=0,clip=true]{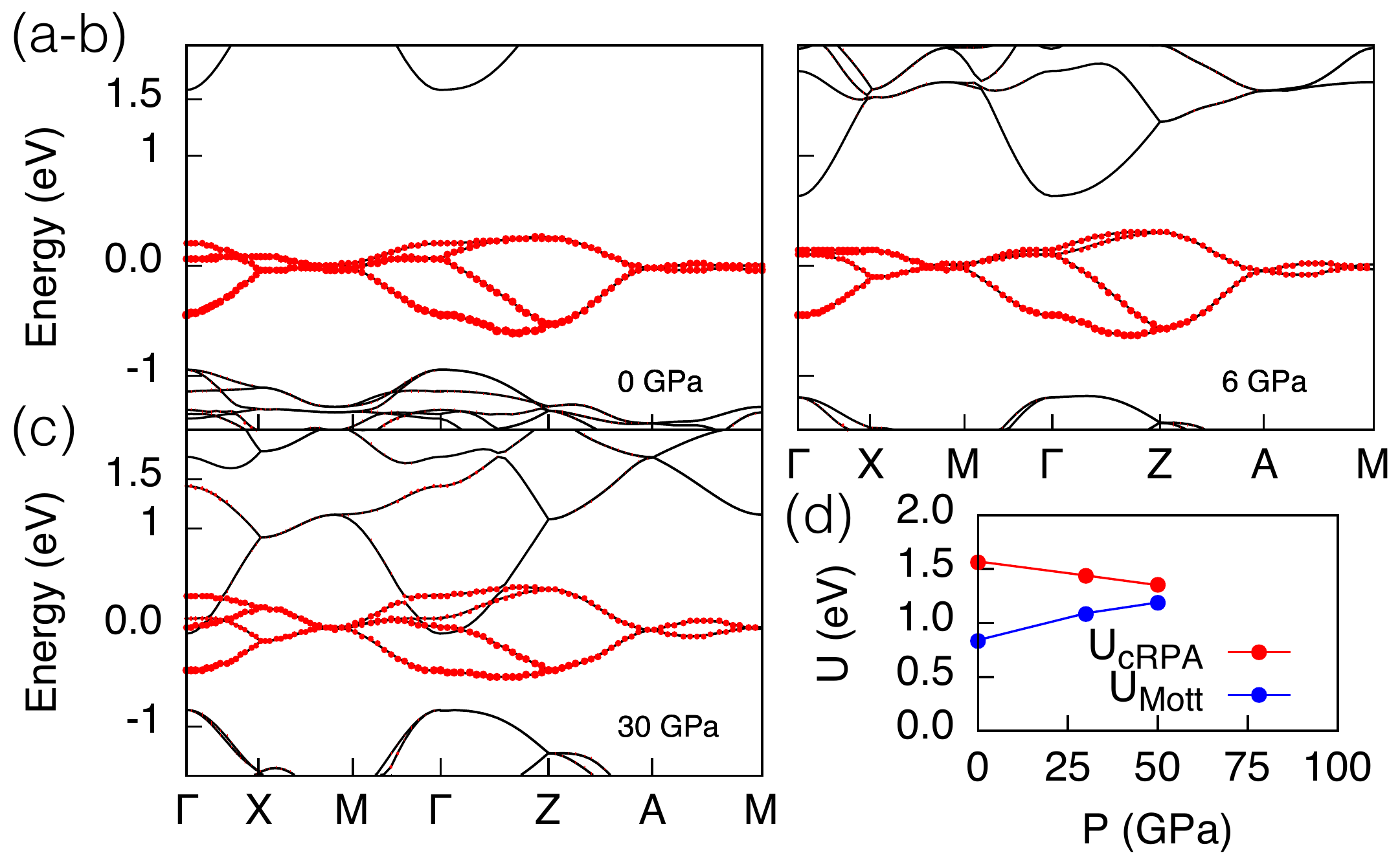}
 \caption{(Color online) (a-c) Evolution of the DFT band structure for $P=0$ (ambient
   pressure), 6, and 30 GPa.
 Red symbols refer to the $d_{x^2-y^2}$ character. (d) $U_{\text{cRPA}}$ and $U_{\text{Mott}}$
 as a function of pressure. It is suggestive to expect an insultar to
 metal transition around $P=50-60$ GPa.}
 \label{fig4}
 \end{figure}
 
{\it Magnetic order and hydrostatic pressure.} At $U_{\text{cRPA}}$,
allowing for magnetic ordering, the DFT+DMFT ground state of {\cubbio} is not
paramagnetic, but displays intra unit cell magnetic order. As a consequence, the
double Dirac cone is absent due to the breaking of time-reversal
symmetry. This finding holds for finite temperature $T$, below the
N\'eel temperature $T_{\text{N}}$. For $T>T_{\text{N}}$, while there
is no long-range magnetic order, the broadening is too strong to
resolve detailed dispersive features. In Tab.~\ref{tab2}, we report the
DFT+U energies at $T=0$ for different magnetic
configurations. 
Mapping these energies on a Heisenberg model
and performing classical Monte Carlo simulations,
we estimate a N\'{e}el temperature of 56 K for U = 1.58 eV, which compares
reasonably well with the experimental value of ~50 K \cite{Supp,GarciaMunoz,Janson}.
The AFM-C type configuration, characterized by a ferromagnetic
alignment of the chains along the $c$-direction, is found to be
preferred. Independent of the considered interaction
strength, however, it is always
energetically close to an AFM-G type phase, where the magnetic moments are antiferromagnetically 
aligned along all three directions.
This is the consequence of a competitive balance between $t^A_1$ and
$t^{AB}_{1d}$ transfer integrals (Tab.~\ref{tab1}). This tuned
frustration is relieved by pressure, favoring $t^A_1$ over
$t^{AB}_{1d}$. It explains why we
observe a change to AFM-G type magnetic order as a function of pressure (see Tab. \ref{tab2}).

Along with the change of the magnetic order, pressure has a profound impact
on the electronic structure. As illustrated in Fig. \ref{fig4}(a-c), the bands above
and below the $d_{x^2-y^2}$ manifold get closer to the Fermi level. This results
in a more efficient screening of the Coulomb interaction for the
target $d_{x^2-y^2}$ states. Fig. \ref{fig4}(d) and TABLE \ref{tab1}
report the corresponding reduction of $U_{\text{cRPA}}$. In addition, the slight increase of the $d_{x^2-y^2}$ bandwidth ($\sim 10$\%)
leads to a larger critical $U_{\text{Mott}}$ for the Mott transition. Even though $U_{\text{cRPA}}$ remains
larger than $U_{\text{Mott}}$ all the way up to 50 GPa \cite{note_pressure}, the trend towards the restoration
of the metallic phase is evident \cite{noteU}. Assuming the absence of
a structural phase transition~\cite{note_pressure}, this suggests a
transition into a
high-pressure double Dirac metal for $\simeq 60$ GPa.

{\it Conclusion.} Our analysis identifies {\cubbio} as a prototypical
material where crystal structure and orbital character conspire to
give rise to correlated double Dirac fermions close to the Fermi
level. At ambient pressure, the interactions turn out to drive the
system into Mott state along with magnetic intra-cell ordering where
the double Dirac cone is absent. As a function of hydrostatic
pressure, we find that the material could be driven into a metallic
state where the double Dirac features would emerge. For this, our ab
initio calculations indicate a pressure regime of $\approx 60$ GPa,
which is still within the range of experimental pressure cell
transport setups. Our study suggests several routes to realize a double-Dirac metal in
{\cubbio} at low temperatures. For instance, a combined pressure and doping
approach could establish a convenient perturbation of the pristine
material in order to render correlated double Dirac fermions accessible
to experimental investigation.


\begin{acknowledgments}
We are grateful to B.~A.~Bernevig for drawing our attention to
double Dirac fermions, and to C. Felser for detailed information on {\cubbio} crystals. 
We thank M.~Baldini,  J.~Cano, A.~Schnyder, and  A. Toschi for helpful discussions. This work was supported by
DFG-SFB 1170 Tocotronics, ERC-StG-336012-Thomale-TOPOLECTRICS, and NSF PHY-1125915.
The authors acknowledge computational resources from the Leibniz Supercomputing Centre
under the Project-ID pr94vu.
\end{acknowledgments}


\bibliographystyle{prsty}
\bibliography{CuBBiO-lib}

\begin{thebibliography}{10}
\bibitem{note_UcRPA}
The value for $U_{\text{cRPA}}$ is defined as the local minus the
  nearest-neighbor part of the cRPA Coulomb matrix element as shown in Tab. I.
\bibitem{note_pressure}
So far, pressure effects have been hardly investigated in {\cubbio}.
  In~\cite{Zhang}, the occurrence of a structural phase transition might have
  been observed within 20 GPa and 37 GPa, but demands further anaylsis.
\bibitem{noteU}
The most reliable feature is the trend observed as a function of pressure
  rather than the absolute values of the Coulomb interaction strength obtained
  through cRPA. The absolute accuracy of cRPA is challenged by (i) the slight
  tendency to underestimate the Coulomb matrix elements, (ii) the mapping of
  the non-local cRPA Coulomb interaction onto a local Hubbard model, and (iii)
  the absence of all non-local correlations within the DMFT approximation. The
  latter is known to reduce the N\'{e}el temperature of about 30-40\% in 3D
  systems \cite{Rohringer}.
\end{thebibliography}

\pagebreak
\clearpage

\section{DFT+DMFT Computational Details}

Density functional theory calculations were performed
by using the VASP ab-initio simulation package \cite{VASP1} within the
projector-augmented-plane-wave (PAW) method \cite{PAW,VASP2}. The
generalized gradient approximation as parametrized by the PBE GGA
functional for the exchange-correlation potential was used \cite{PBE},
by expanding the Kohn-Sham wavefunctions into plane-waves up to an
energy cut-off of 600 eV. The Brillouin zone has been sampled on a
8$\times$8$\times$12 regular mesh, and when considered, spin-orbit
coupling (SOC) was self-consistently included \cite{SOC_VASP}. The low
energy model has been extracted by projecting onto Cu d$_{x^2-y^2}$-like
maximally localized Wannier functions (MLWF) using the WANNIER90 package
\cite{WANNIER90}. Electron-electron interaction was included within the
framework of dynamical mean field theory (DMFT), by mapping the lattice
problem into an impurity model subject to a self-consistency condition
\cite{DMFT}. We solved the impurity model by continuous-time quantum
Monte Carlo, as implemented in the w2dynamics package \cite{CTQMC,w2dynamics}.
 
\section{Setting up the $d_{x^2-y^2}$ low-energy Hamiltonian}

\subsection{{\it Ab initio} computation of the screened Coulomb interaction}

Our aim is to set up a Hamiltonian that faithfully represents the low-energy electronic degrees of freedom of our material.
In particular, we need to compute the strength of the Coulomb repulsion experienced by charge carriers that reside in states
close to the Fermi level. In a solid, the polarizability of charges screens the bare Coulomb repulsion $v(\svek{r},\svek{r}\pr)=\frac{e^2}{4\pi\epsilon_0}1/\left| \svek{r}-\svek{r}\pr\right|$.
When setting up a Hamiltonian for low-energy excitations, its Coulomb interaction thus needs to be screened by all excitations
that are not included in that Hamiltonian.

A successful approach to disentangle and determine screening effects for the Coulomb interaction is the constrained random phase approximation (cRPA)\cite{PhysRevB.70.195104}.
To be on par with the many-body Hamiltonian used in the main manuscript, we here (i) include all contributions to the charge polarization except for those that are confined (constrained)
to the subspace of Cu-$d_{x^2-y^2}$ bands that constitute the low-energy spectrum, and (ii) express the partially screened Coulomb interaction in a basis of maximally localized Wannier functions\cite{miyake:085122}.

The underlying DFT calculations employ a full potential LMTO method~\cite{fplmto}, a Brillouin zone discretized into 8x8x8 $k$-points, local orbitals for the Bi-$5d$ states, and neglects the spin-orbit coupling
which was shown to be weak (see main manuscript). The cRPA is used as implemented in Ref.~\cite{miyake:085122} with a 4x4x4 $k$-mesh.

At ambient conditions, we find that the bare, i.e.\ unscreened, Coulomb interaction $v(\svek{r},\svek{r}\pr)$ in the Wannier basis $\chi^{\hbox{\tiny W}}_{\svek{R}\alpha}(\svek{r})$
\begin{eqnarray}
		&&V^{\alpha\beta\alpha\pr\beta\pr}_{\svek{R},\svek{R}\pr}=\frac{e^2}{4\pi\epsilon_0} \,\times\\
		&&\quad\int d^3r d^3r\pr \chi^{\hbox{\tiny 
		W}*}_{\svek{R}\alpha}(\svek{r}) \chi^{\hbox{\tiny W}}_{\svek{R}\beta}(\svek{r})\frac{1}{\left|\vek{r}-\vek{r}\pr\right|}
	\chi^{\hbox{\tiny W}*}_{{\svek{R}\pr}\alpha\pr}(\svek{r}\pr) \chi^{\hbox{\tiny W}}_{\svek{R}\pr\beta\pr}(\svek{r}\pr)\nonumber 
	\label{eqV}
\end{eqnarray}
amounts to $V^{\alpha\alpha\alpha\alpha}_{\svek{R},\svek{R}}=12.0$eV
for the on-site intra-orbital component of {\cubbio}.
Using the static cRPA polarizability, this interaction is screened down to $U^{\alpha\alpha\alpha\alpha}_{\svek{R},\svek{R}}=2.1$eV.
Figure \ref{figS1}(a) displays the static ($\omega=0$) values of the density-density components $U^{\alpha\alpha\beta\beta}_{\svek{R},\svek{R}}$
within the unit-cell $\svek{R}=0$. In our case orbital-offdiagonal ($\alpha\ne\beta$) components refer to the interaction
between $d_{x^2-y^2}$-derived orbitals of the four equivalent Cu-atoms within the unit-cell. We see that while the decay of the 
screened interaction with distance is still algebraic (in contrast to a Yukawa-type of potential from e.g., the Thomas-Fermi theory of screening), 
the nearest neighbor interaction is already reduced to only 0.52eV, i.e.\ a factor of four smaller than the on-site repulsion.
This finding justifies the use of a Hubbard Hamiltonian in which only intra-atomic interactions are retained.
Note, however, that the on-site Hubbard $U$  is the energetic price for two electrons to be simultaneously on the same atomic site
{\it relative} to occupying two different sites. It is apparent that any non-local interaction will reduce the {\it effective} on-site
interaction in a model that takes into account only local interactions. Indeed, in can be shown, that to first approximation, the effective on-site interaction
is simply given by $U^{eff}=U^{on-site}-U^{nearest-neighbor}$\cite{2015arXiv150807466S}. Therefore, in our case $U^{eff}=1.58eV$.
For the one-band/Cu Hubbard model, the gap in the Mott phase will be roughly given by $U^{eff}$.
Given that the experimental (optical) gap has been estimated to be 1.3-1.8eV\cite{Abdulkarem20111443,doi:10.1021/jp0725533,doi:10.1021/jp210130v},
our Coulomb interaction thus seems very reasonable.

\subsection{Many-body perturbation theory: renormalization of ligand states and exchange self-energies}

In the preceding section we computed the screened Coulomb interaction on the basis of the DFT eigenvalues and eigenfunctions and set up
the effective Hubbard $U$ to be used in dynamical mean-field (DMFT) calculations for the Cu-$d_{x^2-y^2}$ orbitals.

\begin{figure*}[!t]
 \centering
 \includegraphics[width=\columnwidth,angle=0,clip=true]{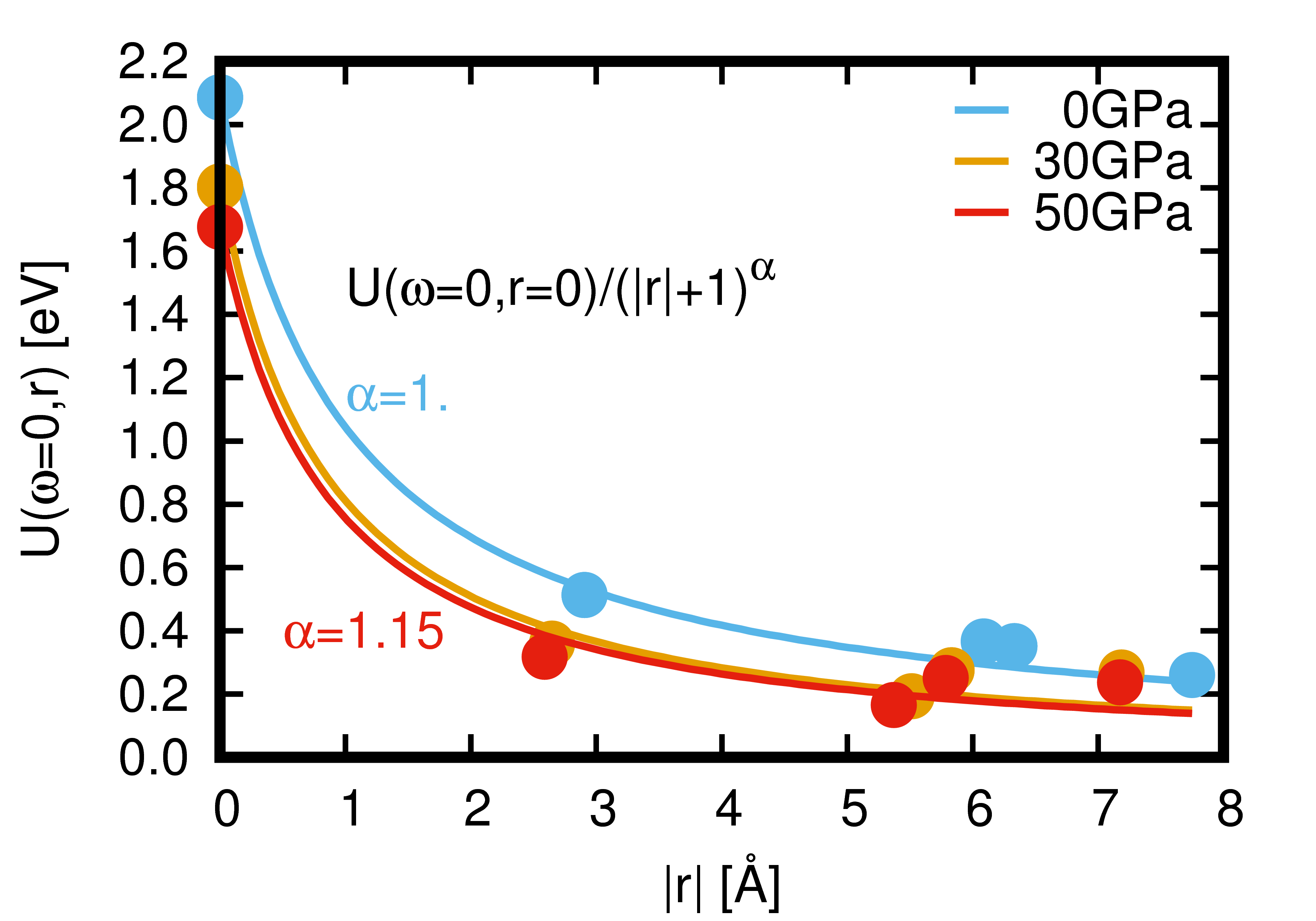}
 \includegraphics[width=\columnwidth,angle=0,clip=true]{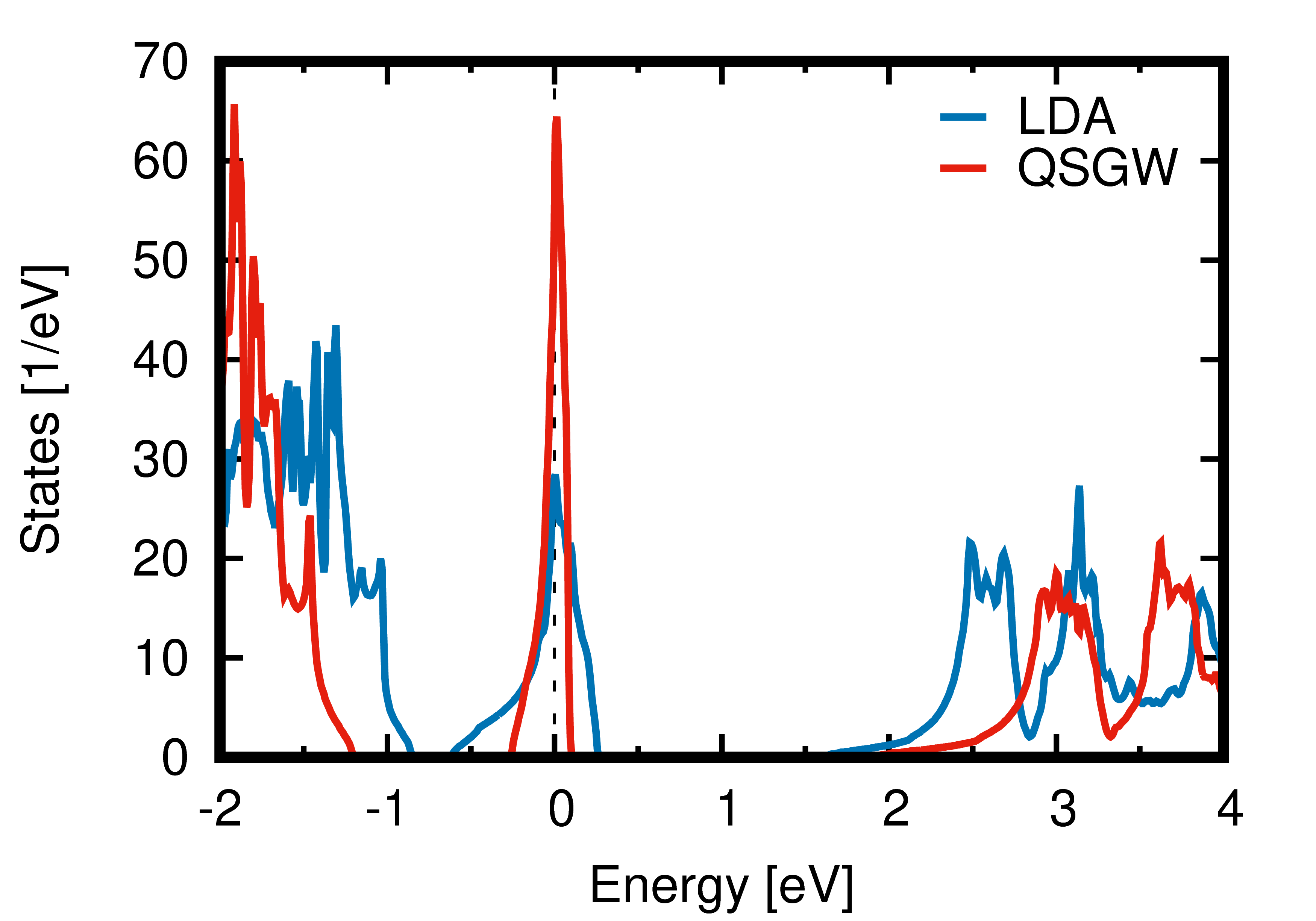} 

 \includegraphics[width=\columnwidth,angle=0,clip=true]{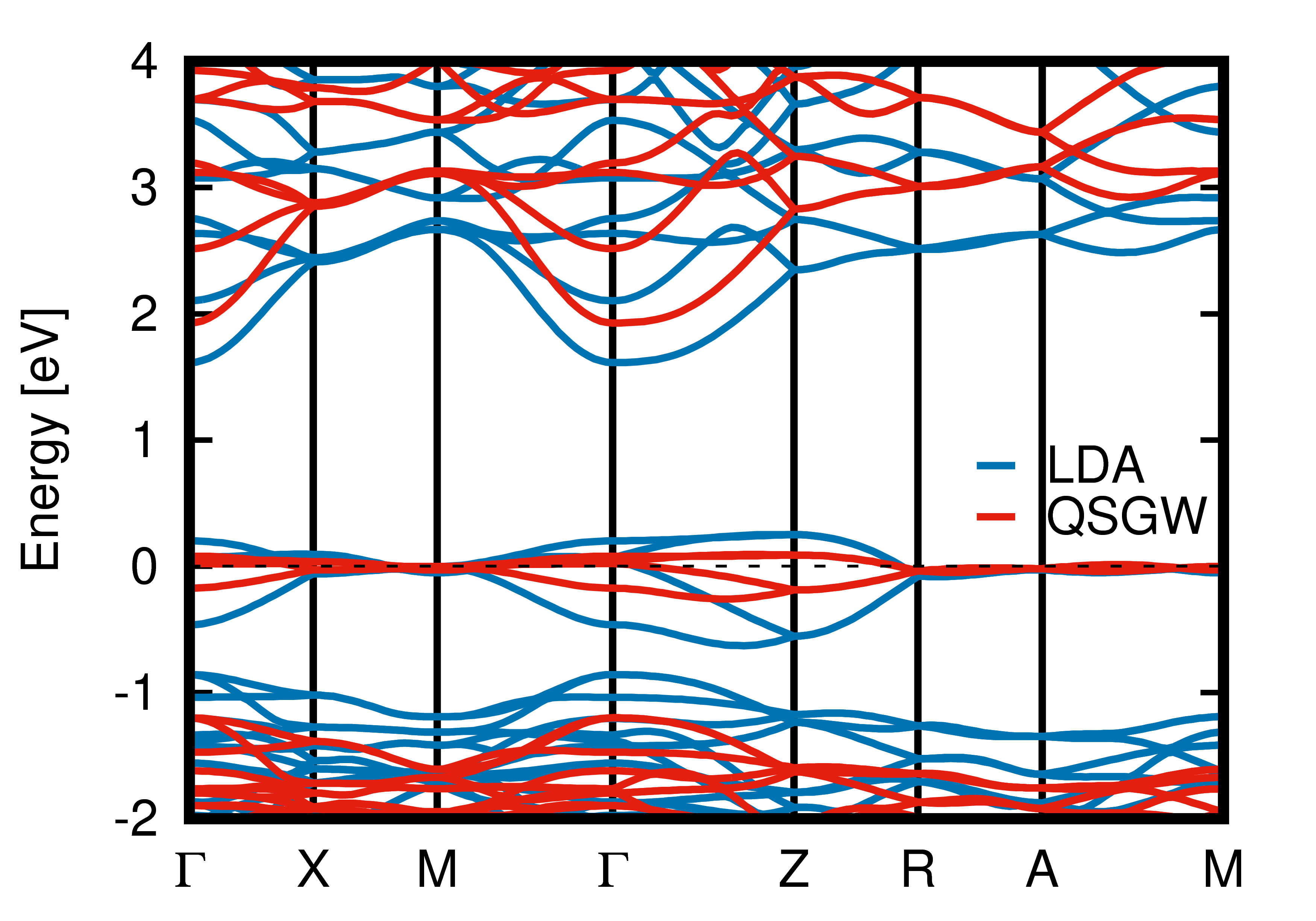}
 \includegraphics[width=\columnwidth,angle=0,clip=true]{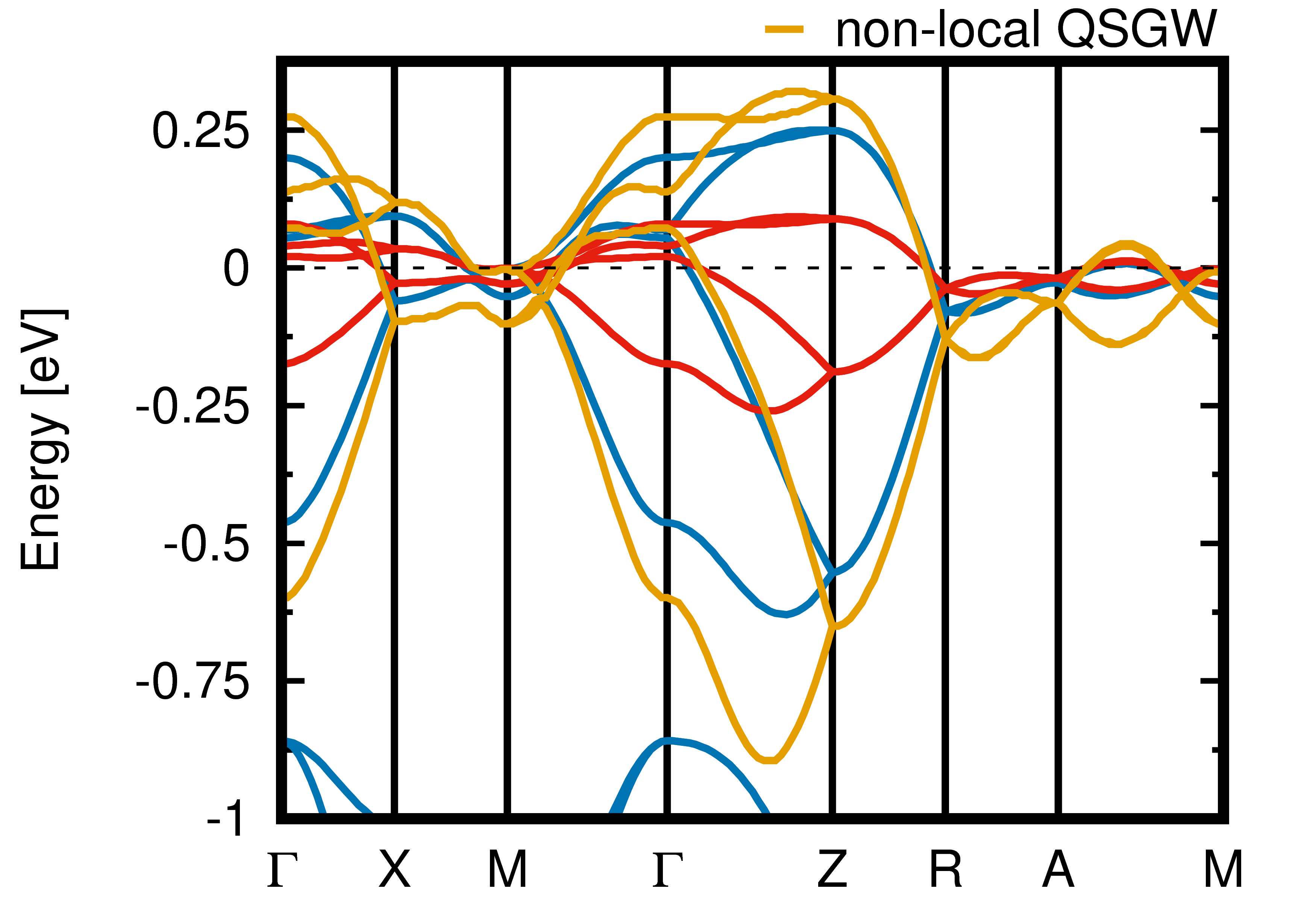}

 \caption{(a) cRPA results for the static Coulomb repulsion within the CuBi$_2$O$_4$ unit-cell.
(b-d) QS{\it GW} calculations for {\cubbio} in comparison with LDA results: (b) density of states, (c,d) band-structures
along the same $k$-path as in the main-manuscript. See text for details.
} \label{figS1}
 \end{figure*}

While this procedure gave a reasonable estimate of the Hubbard $U$, let us point out possible short-comings of this approach.
To this end, we perform electronic structure calculations beyond DFT, using the quasi-particle self-consistent (QS){\it GW}\cite{PhysRevLett.93.126406} approach.
The results are shown Figure \ref{figS1}(b-d) and discussed in the following.

\paragraph{(1) exchange self-energies of the Cu-$d_{x^2-y^2}$}
We first focus on the $d_{x^2-y^2}$ orbitals that are contained in our low-energy  many-body Hamiltonian.
The latter is solved using DMFT, which provides a local, on-site, self-energy $\Sigma$. As discussed in the main manuscript the system is Mott insulating: The quasiparticle weight $Z=[\left. 1-\partial_\omega\Re\Sigma(\omega)]^{-1}\right|_{\omega=0}$ vanishes, as the self-energy diverges.

While the perturbative {\it GW} approximation is incapable to describe the Mott insulating nature of {\cubbio}, we nonetheless witness
a considerable correlation-induced narrowing of the Cu-$d_{x^2-y^2}$ dispersion. Within QS{\it GW} we find a local 
quasi-particle weight $Z=0.29$. From this one would, however, expect a bandwidth-narrowing from $\epsilon^{LDA}(\svek{k})$ to $Z\times \epsilon^{LDA}(\svek{k})$. Yet, as is apparent in
Figure \ref{figS1}(d), the QS{\it GW} bandwidth is only about 50\% smaller than in DFT.
The reason for this is exchange contributions in the {\it GW} self-energy that favour the delocalization of
charges and hence an {\it increase} in the bandwidth\cite{jmt_pnict}. This non-local exchange self-energy is included neither in DFT nor DFT+DMFT,
and is a major argument for more advanced electronic structure theories, that e.g., combine many-body perturbation theory with DMFT.
One example for such a theory is QS{\it GW}+DMFT\cite{jmt_sces14}.

In our case, where there is only one orbital per copper site, we can easily incorporate the effects of the non-local self-energies into an effective Hamiltonian beyond DFT
(for details, see Ref.~\cite{jmt_sces14}): The aim is to use the QS{\it GW} self-energy to construct an improved---{\it non-local} self-energy-containing---one-particle Hamiltonian for usage with DMFT, 
that will add a non-perturbative {\it local} self-energy. To achieve this we have to assure that
local renormalizations are not taken into account twice (``double-counted''). Hence we need to remove all local contributions from the {\it GW} self-energy.
Here, this simply amounts to rescaling the QS{\it GW} bands with the inverse of the local {\it GW} quasi-particle weight $1/Z$. The resulting band-structure is shown in Figure \ref{figS1}(d);
it has a bandwidth that is {\it larger} than the DFT one by 27\%, as it includes the non-local self-energy that tends to delocalize charge carriers. 
Again, owing to the one-band nature of our setup, we can alternatively stick to the DFT Hamiltonian and instead  {\it reduce} the value of the Hubbard $U^{eff}$ by 27\%, yielding $1.15$eV.

\paragraph{(2) ligand states.}
Given the known underestimation of gap values in $sp$-semiconductors and problems of ligand states in e.g., oxides, it stands to reason
to question the DFT as viable starting point for screening processes that involve ligand states.
So now we focus in our QS{\it GW} calculation on bands outside the Cu-$d_{x^2-y^2}$ subspace.
We note that there is a finite (upward) downward shift
of (conduction) valence states.
As a consequence the $d_{x^2-y^2}$ dispersion gets further isolated and screening effects are likely to become less effective.
It can thus be expected that performing the cRPA on top of QS{\it GW} will result
in an {\it increased} Hubbard $U$ for the $d_{x^2-y^2}$ orbitals. See also the discussion on CuBi$_2$O$_4$ under pressure below.
This effect thus competes with the reduction of the Hubbard $U$ discussed in the preceding paragraph.
Performing cRPA calculation on top of QSGW, we find
$U^{on-site}=2.9$eV, $U^{nearest-neighbor}=0.63$eV.
Together with the band-width widening from the preceding section,
we thus find an effective interaction
$U^{eff} = (U-Unn)\times(1-0.27) = (2.9-0.6)\times 0.73 = 1.68$eV
which is almost the same as the LDA result of 1.58eV used in the main text.



\subsection{{\cubbio} under pressure: changes in the Hubbard $U$}

Applying pressure to the system will modify both the one-particle dispersion (see Figure 4 in the main text) and the screened Coulomb interaction (Hubbard $U$).
In order to assess whether pressure could drive the system through an insulator-to-metal transition, we here compute the Hubbard $U$ at the
experimentally accessible pressure of 50GPa. 
Quite in general, the impact of pressure onto the Coulomb interaction is two-fold\cite{jmt_wannier,jmt_mno}: 

\paragraph{(1) basis functions.} The Wannier basis in which hopping and interaction parameters
are expressed is modified. Quite, counter-intuitively, the increased extent of maximally localized Wannier functions under compression {\it increases}
the interaction values\cite{jmt_wannier}. In our case, however, this effect is moderate. Indeed the real-space extend (``spread'') of the Wannier functions
increases from 5.42$\AA^2$ at ambient conditions to 6.52$\AA^2$ at 50GPa. 
In fact we find that the bare, i.e., unscreened Coulomb interaction is basically unaffected by pressure with the on-site matrix element changing from $V=12$eV to $V^{50GPa}=11.7$eV.

\paragraph{(2) screening.}
In our case the change in the Hubbard $U$ is caused by changes in the particle-hole transitions that screen the bare interaction.
These transitions are, within RPA, determined by the one-particle spectrum.
As seen in Fig. 4 of the main text, Bi-p orbitals strongly move down in energy and eventually merge into the d$_{x^2-y^2}$
manifold, causing a sizeable increase of the Coulomb interaction screening. 

Using this electronic structure of {\cubbio} under pressure, we find a value of $U=1.68$eV for the on-site matrix element, and an effective $U$ that incorporates
the nearest neighbor interaction of $U^{eff}=1.36$eV.
This value brings us much closer to the interaction-driven insulator-to-metal transition, which happens at $U_c=1.2$eV for the band-structure corresponding to 50GPa.
These findings advocate that
the double Dirac dispersion can be realized in {\cubbio} under realistic pressures.

\begin{table}%
\begin{tabular}{l||l|l|l}
pressure & 0GPa & 30GPa & 50GPa\\
\hline\hline
on-site $U$ & 2.1 & 1.80 & 1.68\\
nearest-neighbour $U^{nn}$ & 0.51 & 0.36 & 0.32\\
$U^{eff}$ & 1.58 & 1.45 & 1.36
\end{tabular}
\caption{cRPA values for the interaction in the basis of maximally localized Wannier functions using the LDA band-structure.}
\label{tabS1}
\end{table}

\subsection{Many-body perturbation theory: momentum-dependence of the self-energy }

 \begin{figure}[!h]
 \centering
 \includegraphics[width=\columnwidth,angle=0,clip=true]{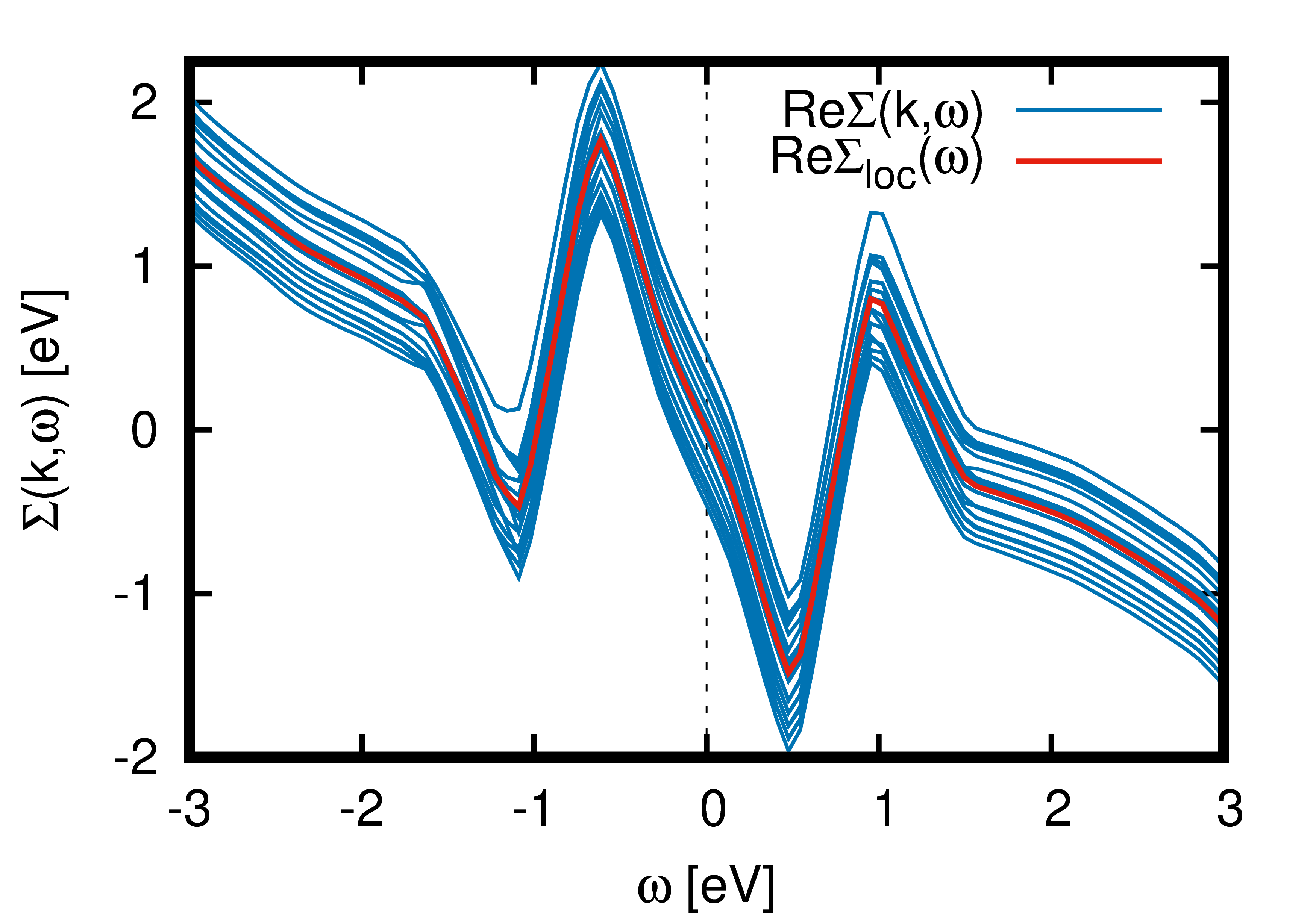}
  \caption{Real parts of the QSGW intra-d$_{x^2-y^2}$ self-energy for all k-points on a 4x4x4 Brillouin-zone mesh (blue), and the local (k-summed) element (red).
	All curves have been arbitrarily shifted by Re$\Sigma_{loc}(\omega=0)$.} \label{figS2}
 \end{figure}
 
 \begin{figure}[!h]
 \centering
 \includegraphics[width=\columnwidth,angle=0,clip=true]{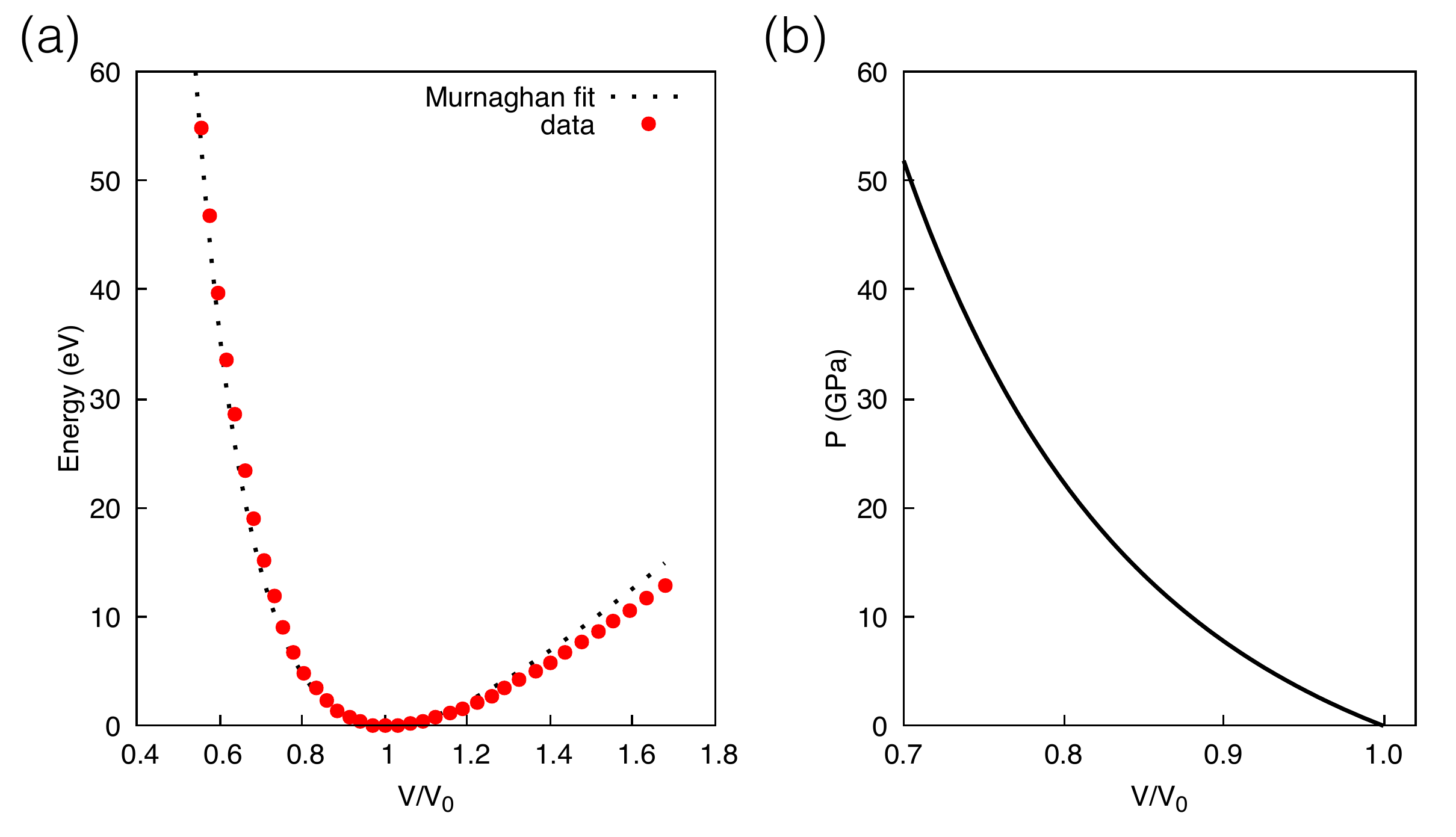}
 \caption{(a) Energy vs. volume curve and (b) pressure vs. volume curve.} 
 \label{figPressure}
 \end{figure} 
 
Fig.~\ref{figS2} shows the intra-d$_{x^2-y^2}$ element of the QSGW self-energy in the maximally localized Wannier basis.
As apparent from the graph, the self-energies for different k-points (different blue curves) are all parallel to each other, i.e.,
they differ only by a constant shift. This means that the dynamics of the self-energy is independent of the momentum. Conversely, this implies
that the non-locality in the self-energy is purely static: $\Sigma(k,\omega)=\Sigma(k)+\Sigma(\omega)$.
This ``space-time separation'' of the self-energy is in-line with recent findings using the non-perturbative dynamical vertex approximation
for the 3D Hubbard model\cite{jmt_dga3d}.

\section{Hydrostatic Pressure}

To estimate the values of pressure we computed the energy versus volume curve, as shown in Fig. \ref{figPressure}(a).
Subsequently this curve was fitted using the following Murnaghan equation of state:

\begin{eqnarray}
E(V) = E_0 + \frac{B_0V}{B_0'}\left[\left(\frac{V_0}{V}\right)^{B_0'}\frac{1}{B_0'-1}+1\right]-\frac{B_0V_0}{B_0'-1}
\end{eqnarray} 

where $E_0$ is the energy minimum, $V_0$ the volume at the minimum, $B_0$ is the bulk modulus and $B_0'$ is its derivative 
with respect to pressure. The obtained parameters were used to produce the $P(V)$ curve

\begin{eqnarray}
P(V) = \frac{B_0}{B_0'}\left[\left(\frac{V_0}{V}\right)^{B_0'}-1\right]
\end{eqnarray} 

shown in Fig. \ref{figPressure}(b).

\begin{figure}
\includegraphics[width=\columnwidth]{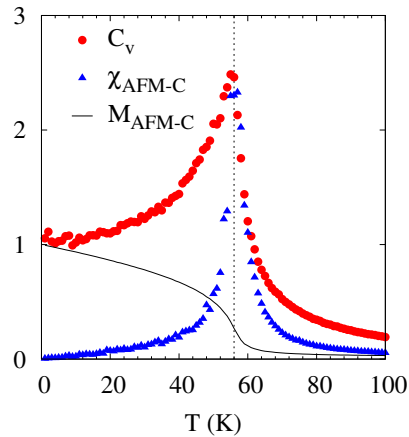}
\caption{Specific heat $C_V$, AFM-C susceptibility $\chi_{AFM-C}$ and order parameter $M_{AFM-C}$ as a function of temperature evaluated from exchange interactions at $U=1.58$eV.}\label{fig_mc}
\end{figure}

\begin{table}[hb]
\begin{tabular}{p{2cm}p{1cm}p{1cm}p{1cm}p{1cm}}
\hline\hline
U (eV) & $J_1$ & $J_2$ & $J_3$ & $J_4$\\
\hline
0.0   & 4.25&5.5&0.07&4.48\\
0.5   & 3.5 &4.81&0.06 &3.72\\
1.58  &	2.38&3.75&0.04 & 2.56\\
2.1   & 2.12&3.38&0.03& 2.29\\
\hline
\end{tabular}
\caption{Estimated exchange interactions (in meV) for different values of the $U$ parameter}\label{tab_J}
\end{table}

\section{N\'eel temperature}

To evaluate the magnetic ordering temperature we map the
total energy of different magnetic phases on a Heisenberg
model
\begin{equation}
H=\frac{1}{2}\sum_{ij} J_{ij}\bm S_i \bm S_j.
\end{equation}
We consider four exchange interactions $J_1,J_2,J_3,J_4$, denoting the
intra-chain interaction between Cu-1(2) and Cu-4(3) along the $c$ axis
($J_1$) and three inter-chain interactions between Cu-1 and Cu-2
($J_2$), Cu-1 and Cu-3 ($J_3$) and between Cu-2 and Cu-4 ($J_4$). Only
the four considered magnetic configurations FM, AFM-C, AFM-G and AFM-A
can be accomodated in the unit cell, which allows to estimate three
independent exchange interactions. Setting the AFM-C energy as the
reference, the Heisenberg parameters can be expressed in terms of total
energy differences as
\begin{eqnarray}
J_1 &=& \frac{1}{8 S^2}\,\left(\Delta E_{FM}-\Delta E_{AFM-A}\right) \nonumber\\
J_2 &=&\frac{1}{16 S^2}\,\Delta E_{FM} \\
J_3+J_4 &=& \frac{1}{8 S^2}\,\left(\Delta E_{FM}+\Delta E_{AFM-G}-\Delta E_{AFM-A}\right) \nonumber
\end{eqnarray}
The ratio $J_3/J_4$ can be then estimated assuming that it is equal to
the ratio $(t_{1,3}/t_{4,2})^2$ between the corresponding transfer
integrals. The estimated exchange interactions for $S=1$ and different
values of the Coulomb parameter $U$ are given in Table \ref{tab_J}.

The Heisenberg model is then used to estimate the N\'eel temperature
from classical Monte Carlo simulations, in order to take spatial
non-local fluctuations beyond mean field, the latter being the only one
included in single-site DMFT.
We used a Metropolis algorithm for
14$\times$14$\times$14 supercells and 10$^7$ Monte Carlo steps for
averages. The diverging beehaviour of the specific heat shown in Fig.
\ref{fig_mc} as a function of temperature allows one to identify the
critical temperature at which the magnetic ordering sets in. In Fig.
\ref{fig_mc} we also show the AFM-C order parameter and its associated
susceptibility, also displaying a divergence at a critical temperature
$T_{N}\simeq 56~$ K.


\end{document}